\documentclass[11pt]{article}
\catcode`\@=11
\@addtoreset{equation}{section}

\baselineskip=\normalbaselineskip

\usepackage{mathrsfs,amsbsy,amssymb,latexsym,amsfonts,amsmath}
\usepackage{graphicx,color}

\allowdisplaybreaks

\begin{document}

\begin{titlepage}
\renewcommand{\thefootnote}{\fnsymbol{footnote}}

\begin{flushright}
\parbox{3.5cm}
{KUNS-2212 \\
RIKEN-TH-159}
\end{flushright}

\vspace*{1.5cm}

\begin{center}
{\Large \bf 
Universal description of viscoelasticity \\ 
with foliation preserving diffeomorphisms}
\end{center}
\vspace{1.5cm}

\centerline{Tatsuo Azeyanagi$^{1}$\footnote{E-mail address: aze@gauge.scphys.kyoto-u.ac.jp},
 Masafumi Fukuma$^{1}$\footnote{E-mail address: 
 fukuma@gauge.scphys.kyoto-u.ac.jp}, 
Hikaru Kawai$^{1,2}$\footnote{E-mail address: 
hkawai@gauge.scphys.kyoto-u.ac.jp}
 and Kentaroh Yoshida$^{1}$\footnote{E-mail address:
 kyoshida@gauge.scphys.kyoto-u.ac.jp}
}

\vspace{2.0cm}

\begin{center}
{\it ${}^{1}$Department of Physics, Kyoto University \\ 
Kyoto 606-8502, Japan\\}
\vspace{0.5cm} 
{\it ${}^{2}$Theoretical Physics Laboratory, Nishina Center, RIKEN\\
Wako, Saitama 351-0198, Japan}
\vspace*{1cm}

\end{center}
\vspace*{1cm}
\begin{abstract}  
A universal description is proposed for generic viscoelastic systems
with a single relaxation time. Foliation preserving diffeomorphisms are introduced as an underlying symmetry 
which naturally interpolates between the two extreme limits 
of elasticity and fluidity. 
The symmetry is found to be powerful enough to determine 
the dynamics in the first order of strains.
\end{abstract}

\thispagestyle{empty}
\setcounter{page}{0}
\end{titlepage}

\newpage
\renewcommand{\thefootnote}{\arabic{footnote}}
\setcounter{footnote}{0}

%%%%%%%%%%%%%%%%%%%%%%%%%%%%%%%%%%%%%%%%%%%%%%%%%%%%%%%%%%%%%%%
\section{Introduction}
%%%%%%%%%%%%%%%%%%%%%%%%%%%%%%%%%%%%%%%%%%%%%%%%%%%%%%%%%%%%%%%
Viscoelasticity is a notion that unifies solids and 
fluids (see \cite{Christensen} 
for example), and is applied to a wide range of materials 
as in rheology \cite{Bingham}.
Viscoelastic materials behave as solids during short intervals 
of time, while they do as ordinary viscous fluids at 
long time scales \cite{LD_fluid, LD_elastic}.

\begin{figure}[htbp]
\begin{center}
\includegraphics[scale=1.5]{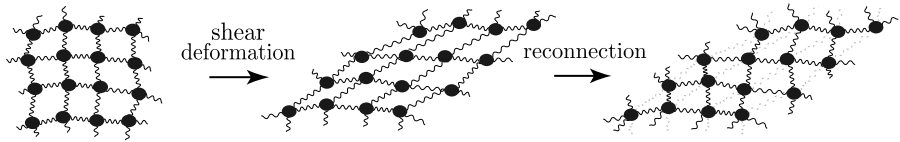}
\parbox[t]{14cm}{\caption{\footnotesize{Shear deformation of viscoelastic material. If we keep the deformation for a long time, 
the bonds are reconnected and then the stress due to the deformation vanishes.}}
\label{reconnection}}
\end{center}
\end{figure}

\begin{figure}[htbp]
\begin{center}
\includegraphics[scale=1.5]{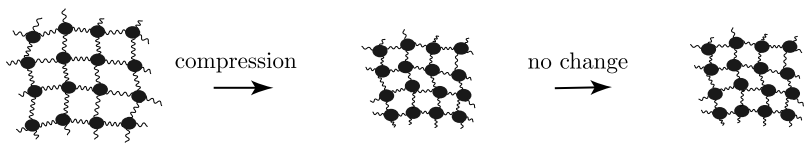}
\parbox[t]{14cm}{\caption{\footnotesize{Compression of viscoelastic material. The material we consider here is elastic for compressions even for 
long time scales, and thus the bulk stress does not undergo relaxation.
}}\label{bulk}}
\end{center}
\end{figure}

In order to get a concrete image for viscoelastic materials, let us imagine something like a chewing gum. 
We can see many particles bonding in it to each other (see Fig.\,\ref{reconnection}). When the material 
is stressed, the bonds produce an elastic force 
and try to make the particles back to the original configuration. That is, the system exhibits elasticity during 
short intervals of time. 
However, if we keep the deformation for a long time, then 
the bonding structure changes to reduce the free energy as 
in Fig.\,\ref{reconnection}, and the shear stress vanishes. 
We further assume that the material is elastic for compressions 
even for long time scales,
so that the bulk stress does not undergo such relaxation (see 
Fig.\,\ref{bulk}). Thus the system exhibits fluidity at long 
intervals of time.

In this paper, we propose a framework in which the 
characteristics of elasticity and fluidity can be both 
dealt with on equal footings. 
We introduce two key ingredients. One is a new variable 
(to be called the intrinsic metric) which represents the 
bonding structures. The other is the 
gauge symmetry of foliation preserving diffeomorphisms 
(FPDs),\footnote{
An interesting application to quantum gravity 
was recently found by Ho$\check{\text{r}}$ava \cite{Horava}.
For mathematical details of FPDs, see also \cite{mathematical}. 
} 
which we find to interpolate between the two extreme limits 
of elasticity and fluidity in a natural manner. 
We show that requiring the invariance 
under FPDs uniquely determines the dynamics of viscoelastic systems 
in the first order of strains.

%%%%%%%%%%%%%%%%%%%%%%%%%%%%%%%%%%%%%%%%%%%%%%%%%%%%%%%%%%%%%%%
\section{Geometrical setup}
\begin{figure}[htbp]
\vspace*{0.5cm}
\begin{center}
\includegraphics[scale=2.0]{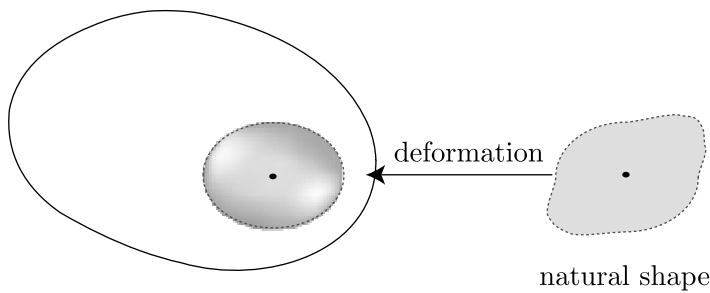}
\parbox[t]{14cm}{\caption{\footnotesize{A sufficiently small region around any point can always be 
regarded as being deformed from the natural shape.}}
\label{chewing}}
\end{center}
\end{figure}

%%%%%%%%%%%%%%%%%%%%%%%%%%%%%%%%%%%%%%%%%%%%%%%%%%%%%%%%%%%%%%%
\label{geometricsetup}
For an elastic material, a sufficiently small region 
around any point can always be regarded as being deformed 
from the configuration with no strains. We call 
the shape before the deformation the natural shape 
(see Fig.\,\ref{chewing}). 
This is a straightforward generalization of the notion 
of the ``natural length" for a spring or a rubber string. 
For a viscoelastic material, we assume that it exhibits the elasticity 
during short time intervals, so that we can define the natural shape around 
a given point at each moment. Since the natural shape is constant in time 
for elastic materials, its time dependence represents the plasticity (i.e.\,non-elasticity) of the material. 
In order to describe the natural shape quantitatively, 
we introduce a new dynamical variable to be called the intrinsic 
metric, and discuss its properties in this section.

\begin{figure}[htbp]
\vspace*{0.5cm}
\begin{center}
\includegraphics[scale=2.0]{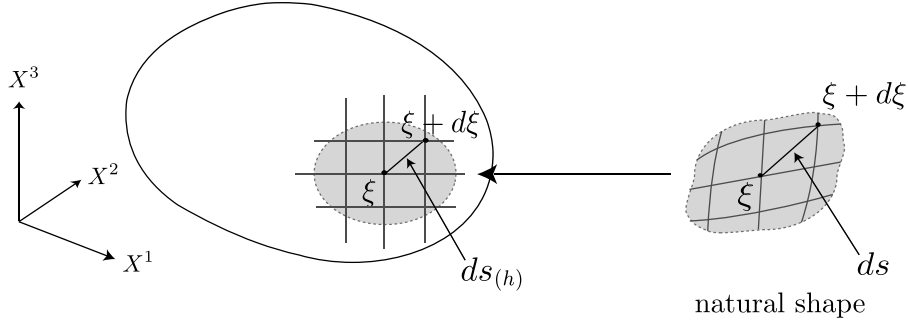}
\parbox[t]{14cm}{\caption{\footnotesize{The intrinsic metric is defined at each time $t$ as 
the distance in the natural shape, $ds^2=g_{ab}(\xi,t)\,d\xi^a d\xi^b$.
The real distance is measured with the induced metric, 
$ds_{(h)}^2 = h_{ab}(\xi, t)\,d\xi^a d\xi^b$.}}
\label{chewing_gum}}
\end{center}
\end{figure}

At each time $t$, we introduce coordinates $\xi=(\xi^a)$ $(a=1,2,3)$ 
arbitrarily on 
the material.\footnote{
Here we assume that the coordinates $\xi$ move smoothly on the material 
as the time $t$ varies.
} 
We then define the intrinsic metric of the material 
(denoted by $g_{ab}(\xi,t)$) such that the distance between two points at $\xi$ and $\xi+d\xi$ 
at fixed time $t$ is given by the distance in the natural shape 
(see Fig.\,\ref{chewing_gum}):
\begin{align}
 ds^2=g_{ab}(\xi,t)\,d\xi^a d\xi^b\equiv 
  \mbox{(length in the natural shape)}^2.
\end{align}

We let $X^i(\xi,t)$ be the spatial 
Cartesian coordinates of the point with $\xi$ at $t$\,. We emphasize  
that the intrinsic metric $g_{ab}(\xi,t)$ is independent of $X^i(\xi,t)$, and 
can differ from the induced metric,\footnote{
 For a function $F(\xi,t)$ we write $\dot{F}=\partial F/\partial t$ 
 and $\partial_a F=\partial F/\partial\xi^a$.}
\begin{align}
 h_{ab}(\xi,t)\equiv \partial_a X^i(\xi,t)\,\partial_b X^i(\xi,t)\,,
\label{intrinsicmet}
\end{align}
which measures the distance in the real three-dimensional space 
(see Fig.\,\ref{chewing_gum}). 
Note that their discrepancies 
\begin{align}
 \varepsilon_{ab}(\xi,t)\equiv \frac{1}{2}\,\bigl(h_{ab}(\xi,t)-g_{ab}(\xi,t)\bigr)
\end{align}
represent the strain tensor of the system. 
In order to see this, let us consider an elastic material. 
We take the coordinates $\xi$ such that they move with the 
atoms and coincide with their Cartesian coordinates in the 
mechanical equilibrium.  
When the material is deformed slightly, 
the atom labeled with $\xi$ 
has the coordinates $X^a(\xi,t)=\xi^a+u^a(\xi,t)$,
where $u^a(\xi,t)$ is the displacement vector, 
and we have
\begin{align}
h_{ab}=\delta_{ab}+\partial_a u_b+\partial_b u_a
+\mathcal{O}(u^2)\,.
\end{align}
Since the intrinsic metric keeps taking the values 
in the mechanical equilibrium, $g_{ab}=\delta_{ab}$\,,
we have that $\varepsilon_{ab}=(1/2)(h_{ab}-g_{ab})
\simeq(1/2)(\partial_a u_b +\partial_b u_a)$. 
This shows that $\varepsilon_{ab}$ is certainly 
the strain tensor for elastic systems. 
In the following discussions, we make the linear approximation 
with respect to $\varepsilon_{ab}$\,. This implies that for the quantities of $\mathcal{O}(\varepsilon_{ab})$ 
one can raise or lower vector indices by either of $g_{ab}$ or $h_{ab}$\,.

We denote by $\rho_0$ the mass density in the absence of strains. 
In this paper, we assume that $\rho_0$ is a constant independent of 
$\xi$ and $t$. 
Then the mass contained in a volume element $d^3\xi=d\xi^1 d\xi^2 d\xi^3$ is given by $\rho_0\sqrt{g(\xi,t)} \,d^3\xi$, 
and 
the mass density in the real three-dimensional space
is given by
\begin{align}
\rho (\xi, t) 
= \frac{\rho_0\sqrt{g(\xi,t)}\,d^3\xi}{\sqrt{h(\xi, t)}\,d^3\xi} 
= \frac{\sqrt{g(\xi,t)}}{\sqrt{h(\xi,t)}}\,\rho_0\,.  
\label{defofrhoh}
\end{align}

\begin{figure}[htbp]
\begin{center}
\includegraphics[scale=1.2]{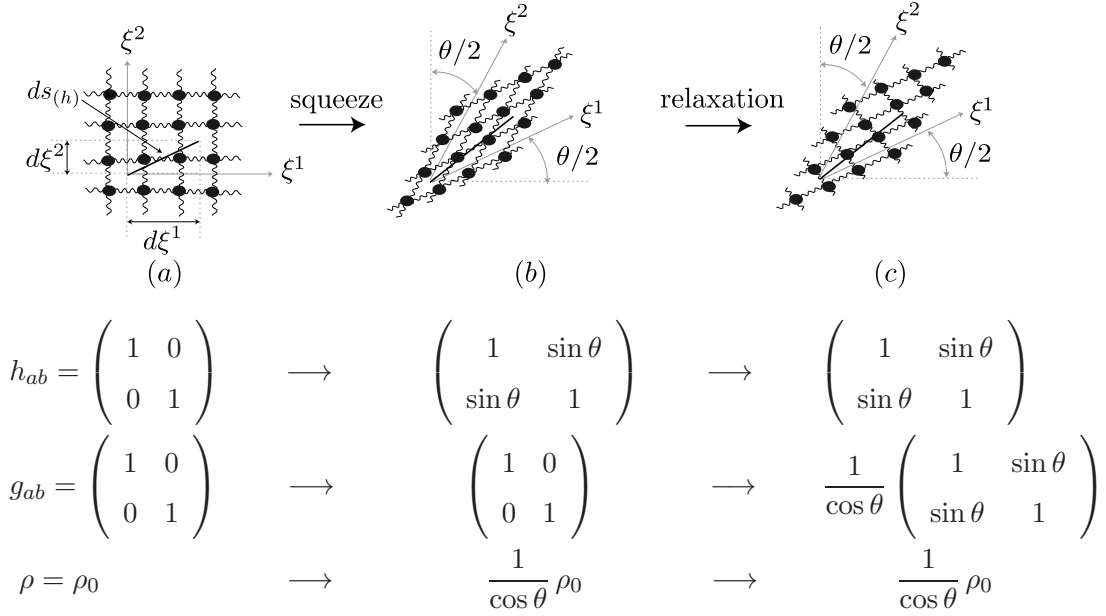}
\begin{align}
&\qquad\quad\,\,h_{ab} = \left(
  \begin{array}{cc}
    1   & 0   \\
    0  & 1   \\
  \end{array}
\right)  
\qquad
\longrightarrow 
\qquad\quad
\left(
  \begin{array}{cc}
    1   &  \sin\theta   \\
    \sin\theta   & 1   \\
  \end{array}
\right) 
\qquad
\longrightarrow 
\qquad
\left(
  \begin{array}{cc}
    1   &  \sin\theta   \\
    \sin\theta   & 1   \\
  \end{array}
\right) 
\nonumber\\
&\qquad\quad\,\,g_{ab} = \left(
  \begin{array}{cc}
     1  &  0  \\
     0  &  1  \\
  \end{array}
\right) 
\qquad\,
\longrightarrow
\qquad\,\quad\quad
 \left(
  \begin{array}{cc}
     1  &  0  \\
     0  &  1  \\
  \end{array}
\right) 
\qquad\quad\,\,\,\,
\longrightarrow
\qquad
 \frac{1}{\cos\theta}\left(
  \begin{array}{cc}
     1  &  \sin\theta  \\
     \sin\theta  &  1  \\
  \end{array}
\right) 
\nonumber \\
&\quad\qquad\,\,\,\,\rho=\rho_0 \nonumber
\qquad\quad\,\qquad\,\,\,\,\,\,
\longrightarrow
\qquad\qquad\quad
\frac{1}{\cos\theta}\,\rho_0\,\,\, 
\qquad\quad\,\,\,\,
\longrightarrow 
\qquad\qquad\,\,\,
\frac{1}{\cos\theta}\,\rho_0 \nonumber
\end{align}
\parbox[t]{14cm}{\caption{\footnotesize{Squeeze with the angle $\theta\,$.
This can be obtained by combining a shear deformation 
and a compression (plus a rotation). }}
\label{example3}}
\end{center}
\end{figure}

As an example, let us consider a squeeze 
deformation of a two-dimensional viscoelastic 
material formed by bonding particles as in Fig.\,\ref{example3}\,. 
The wavy lines stand for the bonds. 
The left figure represents the material before the deformation. 
Here we take $\xi$ to be Cartesian coordinates, and then 
both of the induced and intrinsic metrics have the same form: 
$h_{ab}=g_{ab}=\delta_{ab}$\,. Therefore the mass density $\rho$ 
is equal to $\rho_0$\,. We assume that $\xi$ are attached to the particles, so that they comove with the material under the deformation. 
The middle figure represents the material just after the deformation.
The induced metric and the mass density change according to the 
deformation, while the intrinsic metric does not.
The induced metric is evaluated as 
$ds^2=(d\xi^1)^2+(d\xi^2)^2-2d\xi^1d\xi^2\cos(\pi/2+\theta)$\,
by using the cosine formula.    
The right figure represents the material after a sufficiently long time. 
In the process of relaxation, the induced metric and the mass density 
are preserved, while the intrinsic metric $g_{ab}$ becomes 
proportional to $h_{ab}$\,, \,$g_{ab}=f(\xi,t)\,h_{ab}$\,, as will be discussed around \eqref{gproph}\,. The factor $f(\xi,t)$ can be determined from 
$\rho$ through \eqref{defofrhoh}.

%%%%%%%%%%%%%%%%%%%%%%%%%%%%%%%%%%%%%%%%%%%%%%%%%%%%%%%%%%%%%%%
\section{Foliation preserving diffeomorphisms}
\begin{figure}[htbp]
\vspace*{0.5cm}
\begin{center}
\includegraphics[scale=1.8]{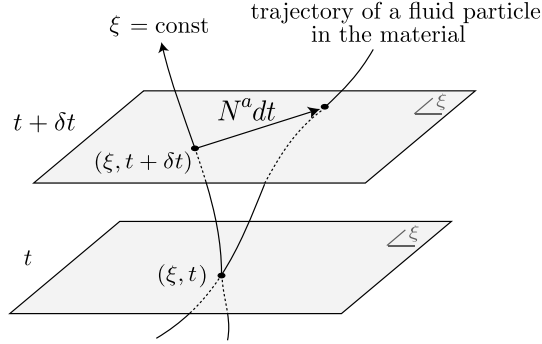}
\parbox[t]{14cm}{\caption{\footnotesize{Definition of $N^a$}} \label{shift}}
\end{center}
\end{figure}

%%%%%%%%%%%%%%%%%%%%%%%%%%%%%%%%%%%%%%%%%%%%%%%%%%%%%%%%%%%%%%%
Since we have introduced a different coordinate system $\xi$ at 
each time, in order to describe the actual motion of a material one 
needs to specify how the fluid particles move relatively to the coordinate system $\xi$. 
This can be realized by introducing a vector field $N^a(\xi,t)$,
with which the fluid particle located at $\xi$ at time $t$ is supposed to move to the position 
$\xi^a+N^a(\xi,t)\delta t$ after the time interval $\delta t$ (see Fig.\,\ref{shift}).

In fact, by using $N^a$, the time derivative along the path of a fluid particle 
(the material derivative) of a scalar quantity $\varphi(\xi,t)$ can be 
expressed as 
\begin{align}
\frac{D\varphi}{Dt} = \dot{\varphi} 
+N^a \partial_a \varphi\,.
\label{scalarcovderiv}
\end{align}
For example, when measured in the real space, 
the velocity $v^i(\xi,t)$ and the acceleration 
$a^i(\xi,t)$ of the fluid particle located at $\xi$ are given by 
\begin{align}
v^i&= \frac{DX^i}{Dt} = \dot{X}^i+N^a\partial_a X^i, 
\label{v^i}\\
a^i&= \frac{Dv^i}{Dt} = \dot{v}^i+N^a\partial_a v^i.    
\label{a^i}
\end{align}

Then we consider the foliation preserving diffeomorphism (FPD):
\begin{align}
t' = t, \qquad \xi'{}^{a}=\xi'{}^{a}(\xi,t).
\label{coordtrans}
\end{align}
We say that a quantity 
$T^{a\cdots}_{~~~b\cdots}(\xi,t)$ is a covariant tensor, if it transforms 
as a three-dimensional tensor at each time: 
\begin{align}
T'\,^{a\cdots}_{~~~b\cdots}(\xi',t)=
\frac{\partial \xi'{}^a}{\partial \xi^c}\cdots\frac{\partial \xi^d}{\partial \xi'{}^b}\cdots
T^{c\cdots}_{~~~d\cdots}(\xi,t)\,, 
\end{align}
where the transition functions 
$\partial \xi'{}^a/\partial \xi^c$ and 
$\partial \xi^d /\partial \xi'{}^b$ should be evaluated at time $t$. 
For example, $X^i$, $v^i$ and $a^i$ are covariant scalars, 
and $g_{ab}$, $h_{ab}$ and $\varepsilon_{ab}$ are 
rank two covariant tensors. 

On the other hand, $N^a$ is not a covariant vector but 
transforms with an inhomogeneous term:\footnote{
This is naturally understood, if one recognizes 
that $N^a$ can be regarded as a gauge field 
that appears when the three-dimensional diffeomorphism 
is gauged in the time direction.}
\begin{align}
N'{}^a(\xi',t) = 
\frac{\partial\xi'{}^a(\xi,t)}{\partial \xi^b}N^b(\xi,t)
+\frac{\partial \xi'{}^a(\xi,t)}{\partial t}\,.
\label{transshift}
\end{align} 
To see this, let us consider a fluid particle in 
two different coordinate systems, and assume that 
it has the coordinates $\xi^a$ and $\xi'{}^a$, respectively, 
at time $t$. Then, by definition, at time $t+\delta t$, 
they become $\xi^a+N(\xi,t)\delta t$ and 
$\xi'{}^a+N'{}^a(\xi',t)\delta t$, respectively. 
Because these are related by the transformation  
\eqref{coordtrans} at $t+\delta t$, we have 
\begin{align}
\xi'{}^a +N'{}^a(\xi',t)\delta t 
&=\xi'{^a}(\xi+N(\xi,t)\delta t,\, t+\delta t) \nonumber \\
&= \xi'{}^a + \frac{\partial\xi'{}^a(\xi,t)}{\partial\xi^b}
N^b(\xi,t)\delta t +\frac{\partial \xi'{}^a(\xi,t)}{\partial t}\delta t,
\end{align}
and thus we obtain \eqref{transshift}.

Note that the time derivative of a covariant tensor 
$T^{a\cdots}_{~~~b\cdots}$ is no longer covariant. 
We, however, can make a covariant tensor by considering 
the derivative along fluid particles as in
\eqref{v^i} and \eqref{a^i}: 
\begin{align}
\frac{D}{Dt}\,T^{a\cdots}_{~~~b\cdots} = \dot{T}^{a\cdots}_{~~~b\cdots}
+{\mathcal L}_{\mathcal N}\, T^{a\cdots}_{~~~b\cdots}\,,
\end{align}
where ${\mathcal L}_{\mathcal N}$ is the Lie derivative 
with respect to the vector field ${\mathcal N}=N^a\partial_a$\,: 
\begin{align}
{\cal L}_{\mathcal N }\, T^{a \cdots}_{~~~b\cdots}= 
N^{c}\partial_{c} T^{a \cdots}_{~~~b\cdots} 
-\partial_{c}N^a T^{c\cdots}_{~~~b\cdots}-\cdots
+\partial_{b}N^c T^{a\cdots}_{~~~c\cdots}+\cdots.
\end{align} 
For example, from $g_{ab}$, we can make a covariant tensor $K_{ab}$ 
to be called the extrinsic curvature, 
\begin{align}
K_{ab} = \frac{1}{2}\,\frac{Dg_{ab}}{Dt}=
\frac{1}{2}\,\bigl(
\dot{g}_{ab}+\partial_a N^c\,g_{cb}
+\partial_b N^c\,g_{ac}
+N^c\,\partial_c g_{ab}\bigr).\label{Kg_ab}
\end{align}
Similarly, we can define the extrinsic curvature $K_{ab}^{(h)}$ 
which corresponds to $h_{ab}$, 
\begin{align}
K_{ab}^{(h)} = \frac{1}{2}\,\frac{Dh_{ab}}{Dt} 
=\frac{1}{2}\,\bigl(
\dot{h}_{ab}+\partial_a N^c\,h_{cb}
+\partial_b N^c\,h_{ac}
+N^c\,\partial_c h_{ab}\bigr).\label{Kh_ab}
\end{align}
Note that  
\begin{align}
 K_{ab}^{(h)} -K_{ab} = \frac{D \varepsilon_{ab}}{Dt}\,.
\label{khabminuskab}
\end{align}

We then introduce the dreibeins $e^i_a(\xi,t)$ and their inverses 
$e^a_i(\xi,t)$ : 
\begin{align}
e^i_a(\xi,t) =  \partial_a X^i(\xi,t)\,, \qquad e^a_i\,e^i_b=\delta^a_b\,.
\label{dreibein} 
\end{align}
From \eqref{v^i} and \eqref{dreibein}, 
we obtain 
\begin{align}
h_{ab}=e_{a}^{i}e_{b}^i\,, \qquad
\partial_{a}e_{b}^i=\partial_b e_a^i\,, \qquad 
\dot{e}_a^i = \partial_a v^i -(\partial_a N^b)e^i_{b}-N^b\partial_a e_b^i\,, \label{usefulid}
\end{align}
and thus  
\begin{align}
\frac{D e_a^i}{Dt} = \partial_{a}v^i. \label{nablaperpe}
\end{align}
By using the dreibeins, the scalars $v^i(\xi,t)$ and $a^i(\xi,t)$ 
can be transformed into vector fields as
\begin{align}
 v^a &\equiv e^a_i v^i,\qquad 
 v_a \equiv e_a^i v^i, \qquad
 a^a \equiv e^a_i a^i, \qquad
 a_a \equiv e_a^i a^i. \label{vaaa} 
\end{align}
We can calculate the acceleration $a_a$ by using 
the Leibniz rule for $D/Dt$ and \eqref{nablaperpe} as
\begin{align}
   a_a&=e_{a}^i\frac{D v^i}{Dt}
     =\frac{D(e_a^iv^i)}{Dt} -v^i\frac{D e_a^i}{Dt}
     =\frac{D v_a}{Dt}-\partial_{a}\left(\frac{1}{2}(v^i)^2\right) \nonumber\\
     &=\dot{v}_a+(\partial_a N^b)v_b
     +N^b\partial_b v_a-\partial_a\left(\frac{1}{2}v^b v_b\right).
\label{3.12}
\end{align}
Furthermore, by using the first of \eqref{usefulid} and 
\eqref{nablaperpe}, we can show 
\begin{align}
K_{ab}^{(h)}=\frac{1}{2}
\left((\partial_a v^i)\, e_b^i+ (\partial_{b} v^i)\, e_{a}^i\right)
=\frac{1}{2}\left(\nabla_a^{(h)}v_b+\nabla_b^{(h)}v_a\right).
\label{Kh_abgeneral}
\end{align}

The set of FPDs forms a gauge symmetry group of the system, 
which can be gauge-fixed arbitrarily 
according to convenience in describing the dynamics of a given system. 
Two of the useful gauge fixings are the followings:

\medskip

\noindent
\underline{\bf (A) comoving frame:} \\
We set $N^a(\xi,t)=0$\,. 
In this frame, $X^i(\xi,t)$ describes 
the motion of the fluid particle attached to the coordinate $\xi$\,, 
and we have 
\begin{align}
v^i(\xi,t)=\dot{X}^i(\xi,t)\,, \qquad a^i(\xi,t)=\ddot{X}^i(\xi,t)\,  \qquad (\text{in the comoving frame}). \label{veloacc}
\end{align}  
This frame is useful for describing the dynamics of elastic materials. 
Note that the acceleration field (\ref{3.12}) is expressed as 
\begin{align}
a_a = \dot{v}_a -\partial_{a}\left(\frac{1}{2}\,v^bv_b\right)\, \qquad (\text{in the comoving frame}). 
\end{align}
The last term gives the inertial force.

\medskip

\noindent
\underline{\bf (B) laboratory frame:} \\
We set $X^a(\xi,t)\equiv \xi^a$.
In this frame, we have $\dot{X}^i(\xi,t)=0$, $e_i^a = \delta_i^a$ and 
$h_{ab} = \delta_{ab}\,$. Then, from \eqref{v^i} and \eqref{vaaa}
we obtain 
\begin{align}
v^a(\xi,t)=N^a(\xi,t)\, \qquad (\text{in the laboratory frame}).
\end{align}
This frame is useful for describing the dynamics of fluid. 
Note that the acceleration field (\ref{3.12}) represents 
the material derivative of the velocity field
\begin{align} 
a_a = \dot{v}_a +v^{b}\,\partial_b v_a  \qquad (\text{in the laboratory frame}). 
\end{align}

%%%%%%%%%%%%%%%%%%%%%%%%%%%%%%%%%%%%%%%%%%%%%%%%%%%%%%%%%%%%%%%
\section{Fundamental equations}
%%%%%%%%%%%%%%%%%%%%%%%%%%%%%%%%%%%%%%%%%%%%%%%%%%%%%%%%%%%%%%%

We are now in a position to write down a set of equations 
which determine the time evolution of $X^i(\xi,t)$, $g_{ab}(\xi,t)$ 
and $N^a(\xi,t)$ up to FPDs. 
The covariance under 
FPDs is found to be powerful enough to 
uniquely determine the dynamics
in the first order of $\varepsilon_{ab}$\,.  

\medskip

\noindent
\underline{\bf $g_{ab}(\xi,t)$}

We first consider the limiting case of elasticity.
Due to our definition of the intrinsic metric, 
$g_{ab}(\xi,t)$ should be constant in time for elastic materials
in the comoving frame ($N^a=0$). 
The FPD-covariant expression for this statement is 
that the extrinsic curvature $K_{ab}$ of (\ref{Kg_ab}) vanishes:
\begin{align}
 K_{ab} = 0\quad \mbox{(elastic limit)}\,.
 \label{elastic_limit}
\end{align}
This implies that the non-vanishing $K_{ab}$ represents the genuinely plastic deformations.

In order to make further discussions, we need to separate the trace part from $K_{ab}$ 
because it vanishes for any materials due to the mass conservation of the system:
\begin{align}
 K(\xi,t)\equiv g^{ab}(\xi,t)K_{ab}(\xi,t) =0\,.
 \label{Feq_mass}
\end{align} 
In fact, contracting (\ref{Kg_ab}) with $g^{ab}$, we obtain
\begin{align}
 K&=\frac{1}{2}\,g^{ab}\bigl(
\dot{g}_{ab}+\partial_a N^c\,g_{cb}
+\partial_b N^c\,g_{ac}
+N^c\,\partial_c g_{ab}\bigr) \nonumber \\
 &=\frac{1}{\sqrt{g}}\,\bigl( (\sqrt{g}\,)^{\mbox{{$\cdot$}}}+\partial_a (\sqrt{g}N^a)\bigr)\nonumber\\
 &=\frac{1}{\rho_{\text{int}}}\,\bigl( \dot{\rho}_{\text{int}}+\partial_a(\rho_{\text{int}} N^a) \bigr),
\end{align}
where $\rho_{\text{int}}(\xi,t)=\rho_0\sqrt{g(\xi,t)}$ is the mass density 
with respect to the intrinsic metric. This indicates that 
the vanishing of $K$ is equivalent to the mass conservation, 
which can be easily seen in the comoving frame where
$N^a=0$ and $\dot{\rho}_{\text{int}}=0$. 

The traceless part of $K_{ab}$ on the other hand 
describes the rate of the shear deformation of the intrinsic metric,  
and thus is expected to be proportional to the traceless part of 
the strain tensor 
\begin{align}
 \tilde{K}_{ab}(\xi,t)
 =\frac{1}{\tau}\,\tilde{\varepsilon}_{ab}(\xi,t)\,,
 \label{Feq_relax}
\end{align}
where $\tau$ is the relaxation time, and 
\begin{align}
\tilde{K}_{ab} &\equiv K_{ab} - \frac{1}{3}Kg_{ab}\,, \\ 
\tilde{\varepsilon}_{ab} &\equiv 
\varepsilon_{ab} -\frac{1}{3}\,(g^{cd}\varepsilon_{cd})\,g_{ab}\,.
\end{align}
We call \eqref{Feq_relax} the rheology equation hereafter. 
After a time interval much longer 
than $\tau$, $\tilde{\varepsilon}_{ab}(\xi,t)$ vanishes and thus 
$g_{ab}(\xi,t)$ becomes proportional to $h_{ab}(\xi,t)$:
\begin{align}
g_{ab}(\xi, t) =f(\xi,t)\,h_{ab}(\xi,t) \quad\quad(f(\xi,t): \text{a scalar function})\,,\label{gproph}  
\end{align}
where $f(\xi,t)$ is determined by the mass conservation 
as in the example in Sec. \ref{geometricsetup}.

The equation (\ref{Feq_relax}) is the simplest and
is expected to be universal. 
It is consistent with (\ref{elastic_limit}) 
because the elastic limit corresponds to $\tau=\infty$\,.
The fluid limit is also realized correctly by taking $\tau=0$\,,
in which we have $\tilde{\varepsilon}_{ab}=0$\,. 
This implies that $g_{ab}$ is proportional to 
$h_{ab}$ as in (\ref{gproph}). 
The only remaining degree of freedom of $g_{ab}$ becomes 
the density of the material, which means that the system 
corresponds to an isotropic fluid.

Note that (\ref{Feq_mass}) and (\ref{Feq_relax}) indicate 
that all the components of $K_{ab}$ are of the order $\varepsilon$. 
Then from \eqref{khabminuskab} 
we find that so are those of $K^{(h)}_{ab}$, 
and we have
\begin{align}
 &\tilde{K}_{ab}^{(h)}-\tilde K_{ab}
 =\frac{D\tilde{\varepsilon}_{ab}^{(h)}}{Dt}+\mathcal{O}(\varepsilon^2)\,,
 \label{khabminuskabtilde}\\
 &K^{(h)}-K=\frac{D\varepsilon^{(h)}}{Dt}+
 \mathcal{O}(\varepsilon^2)\,.
 \label{khminusk} 
\end{align}

\medskip

\noindent
\underline{\bf $X(\xi,t)$}

The dynamics of $X^i(\xi,t)$ should be expressed as Euler's equation 
which is written with the stress tensor $T_{ab}$ as
\begin{align}
 \rho \,a_a= -\nabla^{(h) b} T_{ba}\,,
 \label{Feq_NS}
\end{align}
where $\nabla^{(h)}$ is the covariant derivative 
with respect to $h_{ab}$\,. We show that the leading 
form of the stress tensor in the derivative expansion can be 
determined by the following requirements: 
\begin{itemize}
\item $T_{ab}$ is symmetric and covariant under FPDs.
\item $T_{ab}$ is linear in the strain $\varepsilon_{ab}$ and 
the spatial derivative of the velocity, $\nabla_{a}^{(h)} v_b$.
\end{itemize}

The above requirements 
imply that $T_{ab}$ is a linear combination of the irreducible 
components of $\varepsilon_{ab}$ 
and $K_{ab}^{(h)}=(1/2)(\nabla_{a}^{(h)}v_b+\nabla_{b}^{(h)}v_a)$ 
(see \eqref{Kh_abgeneral}):
\begin{align}
T_{ab} = -2\mu\, \tilde{\varepsilon}_{ab}^{(h)}- 
\frac{1}{\kappa}\,\varepsilon^{(h)}\,h_{ab}-2\gamma\,\tilde{K}_{ab}^{(h)}
-\zeta \,K^{(h)}h_{ab}\,,
\label{stress_general}
\end{align}
where
\begin{align}
\varepsilon^{(h)} &\equiv h^{ab}\varepsilon_{ab}\,,\\
\tilde{\varepsilon}_{ab}^{(h)} &\equiv \varepsilon_{ab}-
\frac{1}{3}\varepsilon^{(h)}h_{ab}\,, \\
K^{(h)} &\equiv h^{ab}K_{ab}^{(h)}\,,\\
\tilde{K}_{ab}^{(h)} &\equiv K_{ab}^{(h)}-\frac{1}{3}K^{(h)}h_{ab}\,.
\end{align}
From the discussions around \eqref{khabminuskabtilde}, 
we find that all the terms in (\ref{stress_general}) 
are of the order of $\varepsilon$. 

In order to see the meaning of the coefficients in (\ref{stress_general}), 
we consider two extreme limits of 
elasticity and fluidity. 
We first consider the elastic limit. 
Since $K_{ab}=0$ in the elastic limit (see \eqref{elastic_limit}), 
the formulas \eqref{khabminuskabtilde} 
and \eqref{khminusk} lead to
\begin{align}
 \tilde{K}_{ab}^{(h)}=\frac{D\tilde{\varepsilon}_{ab}^{(h)}}{Dt}
 \left(1+\mathcal{O}(\varepsilon)\right),\\
 K=\frac{D\varepsilon^{(h)}}{Dt}
 \left(1+\mathcal{O}(\varepsilon)\right),
\end{align}
and thus we have
\begin{align}
 T_{ab}\simeq -2\,\mu\,\tilde{\varepsilon}_{ab}^{(h)}-
  \frac{1}{\kappa}\,\varepsilon^{(h)}h_{ab}
  -2\gamma \frac{D\tilde{\varepsilon}_{ab}^{(h)}}{Dt}
  -\zeta \frac{D\varepsilon^{(h)}}{Dt}h_{ab} 
  \quad (\mbox{elastic limit})\,. 
\end{align}
This indicates that the parameters $\mu$ and $1/\kappa$ 
are the shear and bulk moduli, respectively.
The last two terms express frictions.

On the other hand, the fluid limit is realized 
by considering the case where the time scale $T$ of the variation 
of the shear strain $\tilde\varepsilon_{ab}$,  
\begin{align}
 \frac{D}{Dt}\,\tilde{\varepsilon}_{ab}
 \sim\frac{1}{T}\,\tilde{\varepsilon}_{ab}\,,
 \label{fluid_limit}
\end{align}
is much longer than the relaxation time $\tau$, $T\gg \tau$. 
We then can show that
\begin{align}
 \tilde{K}_{ab}=\tilde{K}_{ab}^{(h)}
 \left(1+\mathcal{O}\left(\frac{\tau}{T}\right)\right)\,,
\end{align}
because the following holds due to \eqref{khabminuskabtilde}:
\begin{align}
 \tilde{K}_{ab}^{(h)}-\tilde{K}_{ab}
 =\frac{D\tilde{\varepsilon}_{ab}^{(h)}}{Dt} 
 \left(1+\mathcal{O}(\varepsilon)\right)
 =\frac{D\tilde{\varepsilon}_{ab}}{Dt}
 \left(1+\mathcal{O}(\varepsilon)\right)
 \sim\frac{1}{T}\,\tilde{\varepsilon}_{ab}=\frac{\tau}{T}\tilde{K}_{ab}\,.
\end{align}
Since 
$\tilde\varepsilon^{(h)}_{ab}
 =\tilde\varepsilon_{ab}\left(1+\mathcal{O}(\varepsilon)\right)=\tau\,\tilde K_{ab}\,\bigl(1+\mathcal{O}(\varepsilon)\bigr)$, 
we can rewrite the stress tensor (\ref{stress_general}) into
\begin{align}
T_{ab} \simeq -2\eta\,\tilde{K}_{ab}^{(h)}
-\zeta \,K^{(h)}h_{ab}-\frac{1}{\kappa}\,\varepsilon^{(h)}\,h_{ab} \quad 
(\mbox{fluid~limit})\,,  
\label{Tab2}
\end{align}
where
\begin{align}
 \eta \equiv \gamma+\mu\,\tau\,.
 \label{eta}
\end{align}
By using \eqref{Kh_abgeneral}, 
each term in (\ref{Tab2}) can be interpreted in terms 
of fluid mechanics if  
we take the laboratory frame ($X^a(\xi,t)=\xi^a$, $v^a=N^a$):
\begin{align}
 h_{ab}&=\delta_{ab}\,,\\
 \tilde{K}^{(h)}_{ab} &= \frac{1}{2}\,(\partial_a v_b + \partial_b v_a )-\frac{1}{3}\,\partial_{c}v_c\,\delta_{ab}\,,\\
 K^{(h)}&=h^{ab}K^{(h)}_{ab}=\partial_c v_c\,,
\end{align}
which indicates that $\eta$ and $\zeta$
represent the shear and bulk viscosities of the fluid, 
respectively.\footnote{
The equation \eqref{eta} shows that the shear viscosity consists 
of two contributions. The first term $\gamma$ reflects 
the friction which the material already has in the elastic 
limit, while the second term $\mu\tau$ represents the stress 
caused by the strain as the material undergoes plastic deformations \cite{LD_elastic}. 
}
The third term in (\ref{Tab2}) can be interpreted as the pressure:
\begin{eqnarray}
p = -\frac{1}{\kappa}\,\varepsilon^{(h)}\,.  
\end{eqnarray}
To see this, we notice that 
$\varepsilon^{(h)}$ ($\simeq (\sqrt{h}-\sqrt{g}\,)/\sqrt{g}$\,)
measures the deviation of the real volume element from that 
of the natural shape,\footnote{
Note that in the linear order of $\varepsilon_{ab}^{(h)}$ we have
\begin{align}
\frac{\sqrt{h}-\sqrt{g}}{\sqrt{g}} = \frac{\sqrt{\det(g_{ab}+2\,\varepsilon_{ab}^{(h)})}-\sqrt{g}}{\sqrt{g}}
\simeq \varepsilon^{(h)}. \nonumber
\end{align}
} 
and that the pressure vanishes in the 
natural shape. Therefore $p$ must be 
proportional to $-\varepsilon^{(h)}$.  
We then find that $\kappa$ agrees with the coefficient of compression
because $\kappa=\rho\,d\rho^{-1}/dp\bigr|_{\rho=\rho_0}$ 
($\rho=\rho_0\sqrt{g}\simeq\rho_0(1-\varepsilon^{(h)})$ and $p=-\kappa^{-1}\varepsilon^{(h)}$).  

The set of equations (\ref{Feq_mass}), (\ref{Feq_relax}), (\ref{Feq_NS}), 
(\ref{stress_general}) and (\ref{eta}) are the fundamental equations 
which govern the dynamics of a given viscoelastic material. 
We can see that macroscopic properties of such materials are characterized 
only by six parameters $(\rho_0,\,\tau,\,\eta,\,\zeta,\,\kappa,\gamma)$ 
from which another parameter $\mu$ is obtained via (\ref{eta}).

%%%%%%%%%%%%%%%%%%%%%%%%%%%%%%%%%%%%%%%%%%%%%%%%%%%%%%%%%%%%%%
\section{Conclusion and Discussion}
%%%%%%%%%%%%%%%%%%%%%%%%%%%%%%%%%%%%%%%%%%%%%%%%%%%%%%%%%%%%%%
We have proposed a set of fundamental equations to
describe generic viscoelastic systems
in a unified way. 
It is expressed as a world
volume theory with the target space
coordinates $X^i(\xi,t)$ and the intrinsic metric 
$g_{ab}(\xi,t)$. FPDs play an important role
in interpolating the two extreme limits of elasticity and fluidity.
We have shown that the covariance under FPDs uniquely 
determines the form of the equations in the first order 
of strains. We thus conclude that the set of equations 
gives a universal description of viscoelastic systems.

Here we have considered viscoelastic systems 
with a single relaxation time. It would be 
interesting to consider a generalization to the case with 
more than one relaxation time in order to describe more 
realistic materials realized in laboratories.

\vspace*{1cm} 

%%%%%%%%%%%%%%%%%%%%%%%%%%%%%%%%%%%%%%%%%%%%%%%%%%%%%%%%%%%%%%%
\begin{center}
\noindent {\bf Acknowledgments}
\end{center}

This work was supported by the Grant-in-Aid for the Global COE program 
``The Next Generation of Physics, Spun from Universality and
Emergence" from the Ministry of Education, Culture, Sports, 
Science and Technology (MEXT) of Japan. 
TA is also supported by the Japan Society 
for the Promotion of Science (JSPS) for Young Scientists 
(No.\,20$\cdot$892). MF and HK are also supported by the 
Grant-in-Aid for Scientific Research No.\,19540288 and 
No.\,18540264, respectively, from MEXT.

%%%%%%%%%%%%%%%%%%%%%%%%%%%%%%%%%%%%%%%%%%%%%%%%%%%%%%%%%%%%%%%


\begin{thebibliography}{99}
%%%%%%%%%%%%%%%%%%%%%%%%%%%%%%%%%%%%%%%%%%%%%%%%%%%%%%%%%%%%%%%
\bibitem{Christensen}
R.~M.~Christensen, 
 ``{\it Theory of Viscoelasticity},''
 Academic Press (1971).

\bibitem{Bingham}
  E.~C.~Bingham,
  ``{\it Fluidity and Plasticity},'' 
  McGraw-Hill (1922).

\bibitem{LD_fluid}
  L.~D.~Landau and E.~M.~Lifshitz,
  ``{\it Fluid Mechanics},'' 
  Butterworth-Heinemann, second edition (1987).

\bibitem{LD_elastic}
  L.~D.~Landau and E.~M.~Lifshitz,
  ``{\it Theory of Elasticity},'' 
  Butterworth-Heinemann, third edition (1986). 
  
\bibitem{Horava}
P.~Ho$\check{\text{r}}$ava,
  ``{\it Membranes at Quantum Criticality},''
  JHEP {\bf 0903} (2009) 020 [arXiv:0812.4287 [hep-th]]. \\ 
P.~Ho$\check{\text{r}}$ava, 
  ``{\it Quantum Gravity at a Lifshitz Point},''
  Phys.\ Rev.\  D {\bf 79} (2009) 084008 
  [arXiv:0901.3775 [hep-th]].  

\bibitem{mathematical}
H.~B.~Lawson Jr., 
``{\it Foliations}," 
Bull.~Amer.~Math.~Soc.~{\bf 80} (1974) 369. \\ 
C.~Godbillon, 
``{ \it Foeuilletages}," Birkh$\ddot{\text{a}}$user (1991). \\ 
I.~Moerdijk and J.~Mr$\check{\text{c}}$um, 
``{\it Introduction to Foliations and Lie Groupoids}," 
Cambridge University Press (2003).


 

\end{thebibliography}
\end{document}